\documentstyle[12pt]{article}

\begin{document}
\title{Effective chiral theory of mesons, coefficients of ChPT,
axial-vector symmetry breaking}
\author{Bing An Li\\
Department of Physics and Astronomy, University of Kentucky\\
Lexington, KY 40506, USA}

\maketitle
\begin{abstract}
A phenomenological successful effective
chiral theory of pseudoscalar, vector, and axial-vector mesons is
introduced. Based on this theory all the 10 coefficients of ChPT are
predicted. It has been found  that a new symmetry breaking-axial-vector
symmetry breaking is responsible for the mass difference between $\rho$
and $a_{1}$ mesons. It is shown that in an EW theory without Higgs
$m_{W}$ and $m_{Z}$ are dynamicallu generated by the combination of the
fermion masses and the axial-vector symmetry breaking: \(m_{W}={1\over\sqrt{2}}
gm_{t}\), \(m_{Z}=\rho m_{W}/cos\theta_{W}\) with $\rho\approx1$, and
\(G_{F}=1/(2\sqrt{2}m^{2}_{t})\). Two gauge fixing terms of W and Z fields
are dynamically generated too.

\end{abstract}

\newpage

\section{Effective chiral theory of mesons}
The chiral perturbation theory is rigorous and phenomenologically
successful in describing the physics of the pseudoscalar mesons at low
energies($E<m_{\rho}$).
Models attempt to deal with the two main frustrations that
the ChPT is limited to pseudoscalar mesons at
low energy and contains many coupling
constants which must be measured.

I have proposed an effective chiral theory of pseudoscalar, vector, and
axial-vector mesons\cite{li}.
It provides a unified study of pseudoscalar, vector, and axial-vector
meson physics at low energies.

The ansatz made in this theory is that
the meson fields are simulated by quark operators.
For example,
\begin{equation}
\rho^{i}_{\mu}=-\frac{1}{g_{\rho}m^{2}_{\rho}}\bar{\psi}\tau_{i}
\gamma_{\mu}\psi.
\end{equation}
The ansatz can be tested. Applying
PCAC, current algebra, and this expression to the
decay $\rho\rightarrow\pi\pi$, under the soft pion approximation
it is derived
\begin{equation}
{1\over 2}f_{\rho\pi\pi}g_{\rho}=1
\end{equation}
which is just the result of the VMD.\\
As a matter of fact, using
fermion operator to simulate a meson field has a long history:
\begin{enumerate}
\item
More than six decades ago, Jordan et al observed
\begin{equation}
{1\over \sqrt{\pi}}\partial_{\mu}\phi=\bar{\psi}\gamma_{5}\gamma_{\mu}
\psi
\end{equation}
in $1+1$ field theory.
\item
A similar relation is proved in the
bosonization of 1+1 field theory.
\item
Use of quark operators to simulate meson fields
has been already exploited in the Nambu-Jona-Lasinio(NJL) model,
a model of four quark interactions.
\item
Quark operators
have been taken as interpolating fields in lattice gauge calculations.
\end{enumerate}

The simulations of the meson fields by quark operators are realized by
a Lagrangian which
is constructed by
chiral symmetry
\begin{eqnarray}
{\cal L}=\bar{\psi}(x)(i\gamma\cdot\partial+\gamma\cdot v
+\gamma\cdot a\gamma_{5}
-mu(x))\psi(x)\nonumber \\
-\bar{\psi(x)}M\psi(x)
+{1\over 2}m^{2}_{0}(\rho^{\mu}_{i}\rho_{\mu i}+K^{*\mu}K^{*}_{\mu}+
\omega^{\mu}\omega_{\mu}+\phi^{\mu}\phi_{\mu}\nonumber \\
+a^{\mu}_{i}a_{\mu i}+K^{\mu}_{1}K_{1\mu}
+f^{\mu}f_{\mu}+f^{\mu}_{1s}f_{1s\mu})
\end{eqnarray}
\(u=exp\{i\gamma_{5}(\tau_{i}\pi_{i}+\lambda_{a}K_{a}+\lambda_{8}
\eta_{8}+\eta_{0})\}\).
To avoide double couting, there are no kinetic terms of mesons, which
are generated dynamically.
Using the least action principle, the relationships between meson fields
and quark operators are found from the Lagrangian. Taking the case of
two flavors as an example, from the least action principle
we obtain
\begin{eqnarray}
{\Pi_{i}\over\sigma}=i(\bar{\psi}
\tau_{i}\gamma_{5}\psi+ix\bar{\psi}\tau_{i}\psi)/(\bar{\psi}
\psi+ix\bar{\psi}\gamma_{5}\psi),\nonumber \\
x=(i\bar{\psi}\gamma_{5}\psi-{\Pi_{i}\over\sigma}
\bar{\psi}\tau_{i}\psi)/
(\bar{\psi}\psi+i{\Pi_{i}\over\sigma}
\bar{\psi}\tau_{i}\gamma_{5}
\psi),\nonumber \\
\rho^{i}_{\mu}=-{1\over m^{2}_{0}}\bar{\psi}\tau_{i}\gamma_{\mu}
\psi,\;\;\;
a^{i}_{\mu}=-{1\over m^{2}_{0}}\bar{\psi}\tau_{i}\gamma_{\mu}
\gamma_{5}\psi,\nonumber \\
\omega_{\mu}=-{1\over m^{2}_{0}}\bar{\psi}\gamma_{\mu}
\psi,\;\;\;
f{\mu}=-{1\over m^{2}_{0}}\bar{\psi}\gamma_{\mu}
\gamma_{5}\psi,
\end{eqnarray}
where \(\sigma+i\gamma_{5}\tau\cdot\Pi=
ue^{-i\eta\gamma_{5}}\)
, \(\sigma=\sqrt{1-\Pi^{2}}\), and \(x=tan\eta\).
The pseudoscalar fields have very complicated quark structures.
Substituting these expressions into the Lagrangian of two flavors,
a Lagrangian of quarks is obtained. It is no longer a theory of four
quarks. This model is different from NJL model.

Using path interal to integrate out the quark fields, the effective Lagrangian
of mesons is derived.
\[{\cal L}_{E}=lndet{\cal D},\]
\[{\cal L}_{re}={1\over2}lndet({\cal D}^{\dag}{\cal D}),\]
\[{\cal L}_{im}={1\over2}lndet({\cal D}/{\cal D}^{\dag}).\]

The question is
whether the masses, decay widths, and interactions
of mesons can be described by the quark operators
correctly. Taking masses as an example,
\begin{enumerate}
\item Pseudoscalar mesons
\begin{eqnarray}
\lefteqn{m^{2}_{\pi^{\pm}}={4\over f^{2}_{\pi^{\pm}}}\{-{1\over3}
<\bar{\psi}\psi>(m_{u}+m_{d})-{F^{2}\over4}(m_{u}+m_{d})^{2}
\},}\nonumber \\
&&m^{2}_{\pi^{0}}={4\over f^{2}_{\pi^{0}}}\{-{1\over3}
<\bar{\psi}\psi>(m_{u}+m_{d})-{F^{2}\over2}(m^{2}_{u}+m^{2}_{d})
\},\nonumber \\
&&m^{2}_{K^{+}}={4\over f^{2}_{K^{+}}}\{-{1\over3}
<\bar{\psi}\psi>(m_{u}+m_{s})-{F^{2}\over4}(m_{u}+m_{s})^{2}
\},\nonumber \\
&&m^{2}_{K^{0}}={4\over f^{2}_{K^{0}}}\{-{1\over3}
<\bar{\psi}\psi>(m_{d}+m_{s})-{F^{2}\over4}(m_{d}+m_{s})^{2}
\},\nonumber \\
&&m^{2}_{\eta_{8}}={4\over f^{2}_{\eta_{8}}}\{-{1\over3}
<\bar{\psi}\psi>{1\over3}(m_{u}+m_{d}+4m_{s})\nonumber \\
&&-{F^{2}\over6}(m^{2}_{u}+m^{2}_{d}+4m^{2}_{s})\},
\end{eqnarray}
The formulas at the first order in quark masses are Gell-Mann, Oakes, and Renner
chiral perturbation theory.
\item Vector mesons
\[m^{2}_{\rho}=m^{2}_{\omega}=6m^{2}\cite{li1}.\]
\item Axial-vector mesons
\[(1-\frac{1}{2\pi^{2}g^{2}})m^{2}_{a}=6m^{2}+m^{2}_{\rho},\]
\[(1-\frac{1}{2\pi^{2}g^{2}})m^{2}_{f}=m^{2}_{\rho}+m^{2}_{\omega},\]
\end{enumerate}
The masses of vector originate in dynamical
chiral symmetry breaking. This is the reason why they are much heavier than
pseudoscalars. The masses of the axial-vector mesons are resulted by a new
symmetry breaking-axial-vector symmetry breaking.

The widths are
\[\Gamma_{\rho}=142MeV(Exp. 150MeV),\]
\[\Gamma_{a}=386MeV(exp. \sim 400MeV),\]
\[\Gamma_{\omega}=7.7MeV(exp. 7.49MeV),\]
\[\Gamma_{f\rightarrow\rho\pi\pi}=
6.01MeV(exp. 6.96(1\pm0.33)MeV)\]

This theory has following features:
\begin{enumerate}
\item The theory is chiral symmetric in the limit of $m_{q}\rightarrow 0$.
The theory has dynamically chiral symmetry breaking(m),
\item VMD is a natural result
\[{e\over f_{v}}\{-{1\over2}F^{\mu\nu}(\partial_{\mu}\rho_{\nu}-
\partial_{\nu}\rho_{\mu})+A^{\mu}j^{\mu}\}.\]
\item Axial-vector currents are bosonized
\[-{g_{W}\over4f_{a}}{1\over f_{a}}\{-{1\over2}F^{i\mu\nu}(\partial_{\mu}
a^{i}_{\nu}-\partial_{\nu}a^{i}_{\mu})+A^{i\mu}j^{i}_{\mu}\}\]
\[-{g_{W}\over4}\Delta m^{2}f_{a}A^{i}_{\mu}a^{i\mu}
-{g_{W}\over4}f_{\pi}A^{i\mu}\partial_{\mu}\pi^{i},\]
Axial-vector symmetry breakin is taken part in.
\item The Wess-Zumino-Witten anomalous action is the leading term of the
imaginary part of the effective Lagrangian,
\item Weinberg's first sum rule is satisfied analytically,
\item The constituent quark mass is introduced as m,
\item Theoretical results of the masses and strong, E$\&$M,
and weak decay widths of mesons agree well with data,
\item The form factors of pion, ${\pi}_{l3}$, $K_{l3}$, $\pi\rightarrow
e\gamma\nu$, and K$\rightarrow e\gamma\nu$ are obtained and agree with data.
For example,
\[<r^{2}>_{\pi}=0.445fm^{2},\]
\[Exp.=0.44\pm0.01fm^{2},\]
\[\rho-pole=
0.39fm^{2}.\]
\item The theory has been applied to $\tau$ mesonic decays\cite{li3}
successfully. For
example,
\[B(\tau\rightarrow\eta3\pi\nu)=3.4\times10^{-4}, \]
\[Exp.=(4.1\pm0.7\pm0.7)
\times10^{-4},\]
\[ChPT=1.2\times10^{-6}.\]
\item $\pi\pi$ and $\pi$K scatterings are studied. Theory agrees with data,
\item The parameters of this theory are: m(quark condensate), g(universal
coupling constant), and three current quark masses,
\item Large $N_{C}$ expansion is natural in this theory. All loop diagrams
of mesons are at higher orders in $N_{C}$ expansion. So far all calculations
are done at the three level,
\item A cut-off has been determined to be 1.6GeV. All the masses of mesons
are below the cut-off. The theory is self consistent,
\end{enumerate}
\section{Coefficients of chiral perturbation theory}
Chiral perturbation theory is the low energy limit of any successful effective
meson theory
\begin{eqnarray}
\lefteqn{{\cal L}={f^{2}_{\pi}\over16}TrD_{\mu}UD^{\mu}U^{\dag}+
{f^{2}_{\pi}\over16}Tr\chi(U+U^{\dag})}\nonumber \\
&&+L_{1}[Tr(D_{\mu}UD^{\mu}U^{\dag})]^{2}+L_{2}(TrD_{\mu}UD_{\nu}U^
{\dag})^{2}\nonumber \\
&&+L_{3}Tr(D_{\mu}UD^{\mu}U^{\dag})^{2}+L_{4}Tr(D_{\mu}U
D^{\mu}U^{\dag})Tr\chi(U+U^{\dag})\nonumber \\
&&+L_{5}TrD_{\mu}UD^{\mu}U^{\dag}(\chi U^{\dag}+U\chi)
+L_{6}[Tr\chi(U+U^{\dag})]
^{2}\nonumber \\
&&+L_{7}[Tr\chi(U-U^{\dag})]^{2}+L_{8}
Tr(\chi U\chi U+\chi U^{\dag}\chi U^{\dag})
\nonumber \\
&&-iL_{9}Tr(F^{L}_{\mu\nu}D^{\mu}UD^{\nu}U^{\dag}+F^{R}_{\mu\nu}
D^{\mu}U^{\dag}D^{\nu}U)\nonumber \\
&&+L_{10}Tr(F^{L}_{\mu\nu}UF^{\mu\nu R}U^{\dag}).
\end{eqnarray}
Many models try to
predict the coefficients of ChPT(see Table I)
\begin{table*}[t]
\caption{Coefficients obtained by various models}
\vspace{0.2cm}
\begin{center}
\begin{tabular}{|c|c|c|c|c|c|c|c|} \hline
&Vectors&Quark&&&Nucleon loop&Linear$\sigma$
model&ENJL\\ \hline
$(L_{1}$+${1\over2}L_{3}$)$\times10^{-3}$&-2.1&-.8&2.1&1.1&-.8&-.5&  \\ \hline
$L_{1}\times10^{-3}$ & & & & & & &0.8                               \\ \hline
$L_{2}\times10^{-3}$ &2.1&1.6&1.6&1.8&.8&1.5&1.6  \\ \hline
$L_{3}\times10^{-3}$ & & & & & & &-4.1 \\ \hline
$L_{4}\times10^{-3}$ & & & & & & &0. \\ \hline
$L_{5}\times10^{-3}$ & & & & & & &1.5 \\ \hline
$L_{6}\times10^{-3}$ & & & & & & &0. \\ \hline
$L_{7}\times10^{-3}$ & & & & & & & \\ \hline
$L_{8}\times10^{-3}$ & & & & & & &0.8 \\ \hline
$L_{9}\times10^{-3}$ &7.3&6.3&6.7&6.1&3.3&.9&6.7 \\ \hline
$L_{10}\times10^{-3}$ &-5.8&-3.2&-5.8&-5.2&-1.7&-2.0&-5.5  \\ \hline
\end{tabular}
\end{center}
\end{table*}
\vspace*{3pt}

The effective chiral theory of mesons is used to predict all the 10
coefficients.
\begin{enumerate}
\item The effective L of $\pi\pi$ scattering at low energy
is derived.
Two of the three coefficients are obtained
\begin{eqnarray}
\lefteqn{2(L_{1}+L_{2})+L_{3}={1\over4}{1\over(4\pi)^{2}}(1-{2c\over g})^{4}}
\nonumber \\
&&L_{2}={1\over4}{c^{4}\over g^{2}}+{1\over4}{1\over(4\pi)^{2}}(1-{2c\over
g})^{2}(1-{4c\over g}-{4c^{2}\over g^{2}})\nonumber \\
&&+{1\over8}(1-{2c\over g}){c\over g}\{2gc+{1\over\pi^{2}}(1-{2c\over g})\}
\nonumber \\
&&c=\frac{f^{2}_{\pi}}{2gm^{2}_{\rho}}.
\end{eqnarray}
\item A complete determination of $L_{1,2,3}$ is carried out from the
effective L of $\pi$K scattering
\begin{eqnarray}
\lefteqn{-2L_{1}+L_{2}=0,}\nonumber \\
&&L_{3}=-{3\over16}{2c\over g}(1-{2c\over g})\{2gc+{1\over\pi^{2}}
(1-{2c\over g})\}\nonumber \\
&&-{1\over2}{1\over(4\pi)^{2}}(1-{2c\over g})^{2}(1-{4c\over g}-{8c^{2}\over
g^{2}})-{3\over4}{c^{4}\over g^{2}},\nonumber \\
&&L_{1}={1\over32}{2c\over g}(1-{2c\over g})\{2gc+{1\over\pi^{2}}
(1-{2c\over g})\}\nonumber \\
&&+{1\over8}{1\over(4\pi)^{2}}(1-{2c\over g})^{2}(1-{4c\over g}
-{4c^{2}\over g^{2}
})+{1\over4}{c^{4}\over g^{2}}.
\end{eqnarray}
\item The coefficients $L_{4-8}$ are determined from
the quark mass expansions of
$m^{2}_{\pi}$, $m^{2}_{K}$, $m^{2}_{\eta}$, $f_{\pi}$, $f_{K}$, and
$f_{\eta}$
\begin{eqnarray}
L_{4}=0,\;\;\;\;L_{6}=0,\\
L_{5}=\frac{f^{2}_{\pi0}f}{8mB_{0}},\\
L_{8}=-\frac{F^{2}}{16B_{0}^{2}},\\
3L_{7}+L_{8}=-\frac{F^{2}}{16B_{0}^{2}},\;\;\;L_{7}=0,
\end{eqnarray}
where
\begin{equation}
B_{0}={4\over f^{2}_{\pi0}}(-{1\over3})<\bar{\psi}\psi>.
\end{equation}
$L_{5}$ and $L_{8}$ are written as
\begin{eqnarray}
L_{5}=\frac{1}{32Q}(1-{2c\over g})\{(1-{2c\over g})^{2}
(1-{1\over2\pi^{2}g^{2}})\nonumber \\
-(1-{2c\over g})+{4\over \pi^{2}}Q
(1-{c\over g})\},\\
L_{8}=-\frac{1}{1536g^{2}Q^{2}}(1-{2c\over g})^{2},
\end{eqnarray}
where
\begin{equation}
Q=-{1\over108g^{4}}{1\over m^{3}}<\bar{\psi}\psi>.
\end{equation}
Q is a function of the universal coupling
constant g only. By fitting $\rho\rightarrow e^{+}e^{-}$, g is determined
to be 0.39.
The numerical value of Q is 4.54.
\item $L_{9}$ and $L_{10}$ are determined by
$<r^{2}>_{\pi}$ and the amplitudes of pion radiative decay,
$\pi^{-}\rightarrow e^{-}\gamma\nu$
\begin{eqnarray}
L_{9}={f^{2}_{\pi}\over48}<r^{2}_{\pi}>,\\
L_{9}={1\over32\pi^{2}}{R\over F^{V}}, \\
L_{10}={1\over32\pi^{2}}{F^{A}\over F^{V}}-L_{9},
\end{eqnarray}
\begin{equation}
L_{9}={1\over4}cg+{1\over16\pi^{2}}
\{(1-{2c\over g})^{2}-4\pi^{2}c^{2}\},
\end{equation}
\begin{equation}
L_{10}=-{1\over4}cg+{1\over4}c^{2}-{1\over32\pi^{2}}
(1-{2c\over g})^{2}.
\end{equation}
\end{enumerate}
There are two
parameters: g and $f^{2}_{\pi}/m^{2}_{\rho}$ in all the 10 coefficients,
which have been determined
already.
The numerical values of the 10 coefficients
predicted by present theory are shown in Table II.
\begin{table*}[t]
\caption{Coefficients from this theory}
\vspace{0.2cm}
\begin{center}
\begin{tabular}{|c|c|c|c|c|c|c|c|c|c|} \hline
$10^{3}L_{1}$&$10^{3}L_{2}$&$10^{3}L_{3}$
&$10^{3}L_{4}$&$10^{3}L_{5}$&$10^{3}L_{6}$
&$10^{3}L_{7}$&$10^{3}L_{8}$&$10^{3}L_{9}$
&$10^{3}L_{10}$  \\ \hline
$.9\pm.3$&$1.7\pm.7$&-$4.4\pm 2.5$&$0.\pm.5$&
$2.2\pm.5$&$0.\pm0.3$&-$.4\pm.15$& $1.1\pm.3$&
$7.4\pm.7$&-$6.\pm.7$ \\ \hline
\end{tabular}
\end{center}
\end{table*}

\section{Axial-vector symmetry breaking and W and Z masses}
Pion, $\rho$ and $a_{1}$ are made of u and d quarks. Pions are Goldstone bosons.
is light.
$m_{\rho}$ is resulted in
dynamical chiral symmetry breaking(quark condensate) and is much heavier
than pions. It is well known that $a_{1}$ meson is the chiral partner of
$\rho$ meson. However, $a_{1}$ is much heavier(1.26GeV) than $\rho$ meson
(0.77GeV) is. \\
Why?\\
What kind of symmetry breaking is responsible for the mass differnce between
$\rho$ and $a_{1}$ mesons?\\
There must be a new dynamical symmetry breaking which generates the mass
difference between $a_{1}$ and $\rho$. It has been found out that the
{\bf \large axial-vector coupling} results a new dynamical symmetry breaking
-axial-vector symmetry breaking which leads to
\[(1-{1\over2\pi^{2}g^{2}})m^{2}_{a}=6m^{2}+m^{2}_{\rho}.\]

Can this axial-vector symmetry breaking bring something to the EW theory?\\
A Lagrangian of EW interactions without Higgs is studied\cite{li2}
\begin{eqnarray}
\lefteqn{{\cal L}=
-{1\over4}A^{i}_{\mu\nu}A^{i\mu\nu}-{1\over4}B_{\mu\nu}B^{\mu\nu}
+\bar{q}\{i\gamma\cdot\partial-M\}q}
\nonumber \\
&&+\bar{q}_{L}\{{g\over2}\tau_{i}
\gamma\cdot A^{i}+g'{Y\over2}\gamma\cdot B\}
q_{L}+\bar{q}_{R}g'{Y\over2}\gamma\cdot Bq_{R}\nonumber \\
&&+\bar{l}\{i\gamma\cdot\partial-M_{f}\}l
+\bar{l}_{L}\{{g\over2}
\tau_{i}\gamma\cdot A^{i}-{g'\over2}\gamma\cdot B\}
l_{L}\nonumber \\
&&-\bar{l}_{R}g'\gamma\cdot B l_{R}.
\end{eqnarray}
There are no $m_{W,Z}$ and there are fermion mass terms which break the
charged part of the $SU(2)_{L}\times U(1)$ symmetry. This explicit
symmetry breaking makes charged bosons, W, massive. However, there are still
two U(1) symmetries. Z and photon are massless.
{\bf \large In EW interactions there are axial-vector couplings, therefore,
there
are	
axial-vector symmetry breaking which contribute mass to both W and Z bosons}.
Using path integral to integrate out the fermion fields, the Lagrangian of boson
fields is derived. After multiplicative renormalization following
results are obtained
\begin{enumerate}
\item
\begin{eqnarray}
m^{2}_{W}={1\over2}g^{2}\{m^{2}_{t}+m^{2}_{b}+m^{2}_{c}+m^{2}_{s}
+m^{2}_{u}+m^{2}_{d}\nonumber \\
+m^{2}_{\nu_{e}}+m^{2}_{e}
+m^{2}_{\nu_{\mu}}+m^{2}_{\mu}+m^{2}_{\nu_{\tau}}+m^{2}_{\tau}\}
\end{eqnarray}
\begin{eqnarray}
m_{W}={g\over\sqrt{2}}m_{t}.
\end{eqnarray}
Using the values \(g=0.642\) and \(m_{t}=180\pm12 GeV\), it is found
\begin{equation}
m_{W}=81.71(1\pm0.067) GeV,
\end{equation}
which is in excellent agreement with data $80.33\pm0.15$GeV.
\item
\[G_{F}=\frac{1}{2\sqrt{2}m^{2}_{t}}=0.96\times10^{-5}(1\pm0.13)m^{-2}_{N}.\]
\item
\begin{equation}
m^{2}_{Z}=\rho m^{2}_{W}/cos^{2}\theta_{W},
\end{equation}
\[\rho=(1-{\alpha\over4\pi}f_{4})^{-1},\]
\[f_{4}={1\over3}N_{G}-{2\over3}\sum_{q}f_{3}+{2\over3}\sum_{l}f_{3}.\]
\[f_{3}=
{1\over2}{1\over\sqrt{x}}ln\frac{1+\sqrt{x}}
{1-\sqrt{x}}.\]
\[x=({m^{2}_{2}\over m^{2}_{1}})^{2},\;\;m_{1,2}={1\over2}(m_{t}\pm m_{b}).\]
\[\rho\sim 1,\]
\[m^{2}_{Z}=m^{2}_{W}/cos^{2}\theta_{W}.\]
\item
The propagator of W field is derived
\begin{equation}
\frac{i}{q^{2}-m^{2}_{W}}\{-g_{\mu\nu}+\frac{q_{\mu}q_{\nu}}{q^{2}}
\}-\frac{i}{\xi_{W}q^{2}-m^{2}_{W}}\frac{q_{\mu}q_{\nu}}{q^{2}}.
\end{equation}
Changing the index W to Z,
the propagator of Z boson field is obtained.
\[\xi_{W}=\frac{N_{G}}{(4\pi)^{2}}{g^{2}\over3},\;\;\;
\xi_{Z}=\frac{N_{G}}{(4\pi)^{2}}(1-{\alpha\over4\pi}f_{4})^{-1}
{1\over3}(g^{2}+g'),\]
\[N_{G}=3N_{C}+3.\]
\end{enumerate}

\begin{enumerate}
\item $m_{W,Z}$ are correctly dynamically generated by axial-vector symmetry bre
\item The propagators of W and Z fields have no problems with renormalization.
\end{enumerate}
\section{Conclusions}
\begin{enumerate}
\item The effective chiral theory is phenomenologically successful,
\item All the 10 coefficients of ChPT are predicted,
\item What can we learned about QCD from this theory:
\begin{enumerate}
\item Chiral symmetry,
\item Dynamical chiral symmetry breaking,
\item Large $N_{C}$ expansion,
\item Simulations of meson fields by quark operators are working. It is worth to
do theoretical study on whether the simulations are solutions of QCD at least
in the limit of large $N_{C}$ expansion,
\end{enumerate}
\item The Lagrabgian is not closed. The quark condensate, m, should be related
to gluons,
\item Axial-vector symmetry breaking provides an explanation to the mass
difference between $\rho$ and $a_{1}$,
\item $m_{W,Z}$ are dynamically generated from
the fermion masses and the axial-vector symmetry breaking.
\end{enumerate}

\end{document}